\documentclass[12pt]{article}
																																																																																											\pdfoutput=1																																																																																			 
\usepackage{amsmath,amssymb}
\usepackage{graphicx}
\usepackage[pdftex]{hyperref}
\DeclareGraphicsExtensions{.eps,.bmp,.wmf,.jpeg,.pdf}
\numberwithin{equation}{section}
\def\be{\begin{equation}}
\def\ee{\end{equation}}

\def\bea{\begin{eqnarray}}
\def\eea{\end{eqnarray}}
\title{Late time cosmological scenarios from scalar field with Gauss Bonnet and non-minimal kinetic couplings}
\author{L.N. Granda\thanks{ngranda@univalle.edu.co} \\ {\small\it Departamento de Fisica, Universidad del Valle}\\{\small\it A.A. 25360, Cali, Colombia}} 
\date{}
\begin{document}
\maketitle

\begin{abstract}
\noindent  We study a model of scalar field with kinetic terms non-minimally coupled to to the curvature, and the scalar field coupled to the Gauss Bonnet 4-dimensional invariant. A variety of solutions are found, giving rise to different cosmological scenarios. A unified description of early time matter (radiation) dominance with transitions to late time quintessence and phantom phases have been made. Among others, we found solutions unifying asymptotically the early power law behavior and late time cosmological constant. Solutions of Chaplygin gas and generalized Chaplygin gas cosmologies have also been found.\\ 

\end{abstract}

\section{Introduction}
\noindent 
The late time acceleration of the universe is one of the most important problems of modern cosmology, which is supported by astrophysical data from distant Ia supernovae observations \cite{riess}, \cite{perlmutter}, \cite{kowalski}, \cite{hicken}, cosmic microwave background anisotropy \cite{komatsu}, and large scale galaxy surveys \cite{percival}. The interpretation of astrophysical observations indicates that this accelerated expansion is due to some kind of  negative-pressure form of matter known as dark energy (\cite{copeland}, \cite{sahnii}, \cite{padmanabhan}). The combined analysis of cosmological observations also suggests that the universe is spatially flat, and  consists of about $\sim 1/3$ of dark matter, and $\sim 2/3$ of homogeneously distributed dark energy with negative pressure. The dark energy may consist of cosmological constant, conventionally associated with the energy of the vacuum \cite{peebles}, \cite{padmana1}, or alternatively, could came from a dynamical varying scalar field at late times which also account for the missing energy density in the universe. A widely explored scalar field models are quintessence \cite{RP}, \cite{wett}, \cite{copeland97}, tachyon \cite{pad}, phantom \cite{caldwell}, K-essence \cite{stein3},\cite{chiba}, and dilaton \cite{gasperini} (for a review see \cite{copeland}). The scalar fields are allowed from several theories in particle physics and in multidimensional gravity, like Kaluza-Klein theory, String Theory or Supergravity, in which the scalar field appears in a natural way. Another alternative to the explanation
of the DE is represented by the scalar-tensor theories, which contain a direct coupling of the scalar field to the curvature, providing in principle a mechanism to
evade the coincidence problem, and naturally allowing (in some cases) the crossing of the phantom barrier \cite{peri}, \cite{maeda}. From a pure geometrical point of view, the modified gravity theories, which are generalizations of the general relativity, have been widely considered to describe the early-time inflation and late-time acceleration, without the introduction of any other dark component, and represent an important alternative to explain the dark energy (for review see \cite{sergei11} and references therein).\\
In the present work, we consider a model with non-minimal coupling to gravity \cite{amendola2, capozziello1, capozziello2}, and with additional Gauss Bonnet coupling, specifically we focus in a scalar field model with kinetic term non-minimally coupled to gravity and to itself \cite{granda,granda1,granda2}, with a new term containing the Gauss Bonnet (GB) 4-dimensional invariant coupled to the scalar field, with the coupling given by an arbitrary function of the field. Despite the fact that the GB term is topologically invariant in four dimensions, and hence, by itself does not contribute to the equations of motion, nevertheless it affects cosmological dynamics when it is coupled to a dynamically evolving scalar field. Besides this, if the GB term is coupled to the scalar field through arbitrary function $f(\phi)$, then this is the unique quadratic combination of the Riemann curvature tensor, that does not increase the differential order of the equations of motion (i.e. GB produces only terms which are second derivatives of the metric in the field equations). Therefore, the coupled GB term which preserves the theory ghost free, seems as a natural generalization of the scalar field with non-minimal kinetic coupling to curvature, which under certain condition is a second order scalar tensor theory. 
Both couplings my have origin in the low energy limit of higher dimensional theories. Thus, the kinetic coupling appears as part of the Weyl anomaly in $N=4$ conformal supergravity \cite{tseytlin, odintsov2}, while the GB coupling arises naturally in the leading order of the $\alpha'$ expansion of string theory \cite{callan}, \cite{deser}. Besides that, the kinetic couplings to curvature are also present as quantum corrections to Brans-Dicke theory \cite{elizalde} and in different frames in modified gravity \cite{sergei11}.\\
Some late time cosmological aspects of scalar field model with derivative couplings to curvature have been considered in \cite{sushkov}, \cite{saridakis}, \cite{gao}. On the other hand, the GB invariant coupled to scalar field have been extensively studied. In \cite{sergei12} the GB correction was proposed to study the dynamics of dark energy, where it was found that quintessence or phantom phase may occur in the late time universe. Accelerating cosmologies with GB correction in four and higher dimensions have been discussed in \cite{tsujikawa}, \cite{leith}, \cite{maartens}. The modified GB theory applied to dark energy have been suggested in \cite{sergei14}, and different aspects of the modified GB model applied to late time acceleration, have been considered among others, in \cite{sergei11}, \cite{sergei15}, \cite{carter}, \cite{tretyakov}.\\
All these studies demonstrate that it is quite plausible that some scalar-tensor couplings predicted by the fundamental theory may become important at current, low-curvature
universe. In the present paper we study different cosmological solutions of the model with Gauss Bonnet coupling to the scalar field and kinetic coupling to curvature. Different solutions unifying early time decelerated behavior, late time accelerated quintessence and phantom phases will be studied. In particular, solutions describing the Chaplygin gas and the generalized Chaplygin gas have been considered.
\section{Field Equations}
Let us start with the  action, for scalar field kinetic terms non-minimally coupled to curvature and coupled to Gauss Bonnet (GB) curvature
\be\label{eq1}
\begin{aligned}
S=&\int d^{4}x\sqrt{-g}\Big[\frac{1}{16\pi G} R-\frac{1}{2}\partial_{\mu}\phi\partial^{\mu}\phi-\frac{1}{2} \xi R \left(F_1(\phi)\partial_{\mu}\phi\partial^{\mu}\phi\right) -\\ 
&\frac{1}{2} \eta R_{\mu\nu}\left(F_1(\phi)\partial^{\mu}\phi\partial^{\nu}\phi\right) - V(\phi)+F_2(\phi){\cal G}\Big]+S_m.
\end{aligned}
\ee
\noindent where {\cal G} is the 4-dimensional GB invariant ${\cal G}=R^2-4R_{\mu\nu}R^{\mu\nu}+R_{\mu\nu\rho\sigma}R^{\mu\nu\rho\sigma}$, $S_m$ is the dark matter action which describes a fluid with barotropic equation of state. The dimensionality of the coupling constants $\xi$ and $\eta$ depends on the type of function $F_1(\phi)$, and the coupling $F_2(\phi)$ is dimensionless. Besides the couplings of curvatures with kinetic terms, one may expect that the presence of GB coupling term may be relevant for the explanation of dark energy phenomena.\\
Taking the variation of action (\ref{eq1}) with respect to the metric, we obtain a general expression of the form 
\be\label{eq2}
R_{\mu\nu}-\frac{1}{2}g_{\mu\nu}R=\kappa^2\left[T_{\mu\nu}^m+T_{\mu\nu}\right]
\ee
where $\kappa^2=8\pi G$, $T_{\mu\nu}^m$ is the usual energy-momentum tensor for matter component, the tensor $T_{\mu\nu}$ represents the variation of the terms which depend on the scalar field $\phi$ and can be written as
\be\label{eq3}
T_{\mu\nu}=T_{\mu\nu}^{\phi}+T_{\mu\nu}^{\xi}+T_{\mu\nu}^{\eta}+T_{\mu\nu}^{GB}
\ee
where $T_{\mu\nu}^{\phi}$,  correspond to the variations of the standard minimally coupled terms, $T_{\mu\nu}^{\xi}$, $T_{\mu\nu}^{\eta}$ come from the $\xi$ and $\eta$ kinetic couplings respectively, and $T_{\mu\nu}^{GB}$ comes from the variation of the coupling with GB. 
Due to the kinetic coupling with curvature and the GB coupling, the quantities derived from this energy-momentum tensors will be considered as effective ones. The variations are given by
\be\label{eq4}
T_{\mu\nu}^{\phi}=\nabla_{\mu}\phi\nabla_{\nu}\phi-\frac{1}{2}g_{\mu\nu}\nabla_{\lambda}\phi\nabla^{\lambda}\phi
-g_{\mu\nu}V(\phi)
\ee
\be\label{eq5}
\begin{aligned}
T_{\mu\nu}^{\xi}=&\xi\Big[\left(R_{\mu\nu}-\frac{1}{2}g_{\mu\nu}R\right)\left(F_1(\phi)\nabla_{\lambda}\phi\nabla^{\lambda}\phi\right)+g_{\mu\nu}\nabla_{\lambda}\nabla^{\lambda}\left(F_1(\phi)\nabla_{\gamma}\phi\nabla^{\gamma}\phi\right)\\
&-\frac{1}{2}(\nabla_{\mu}\nabla_{\nu}+\nabla_{\nu}\nabla_{\mu})\left(F_1(\phi)\nabla_{\lambda}\phi\nabla^{\lambda}\phi\right)+R\left(F_1(\phi)\nabla_{\mu}\phi\nabla_{\nu}\phi\right)\Big]
\end{aligned}
\ee
\be\label{eq6}
\begin{aligned}
T_{\mu\nu}^{\eta}=&\eta\Big[F_1(\phi)\left(R_{\mu\lambda}\nabla^{\lambda}\phi\nabla_{\nu}\phi+R_{\nu\lambda}\nabla^{\lambda}\phi\nabla_{\mu}\phi\right)-\frac{1}{2}g_{\mu\nu}R_{\lambda\gamma}\left(F_1(\phi)\nabla^{\lambda}\phi\nabla^{\gamma}\phi\right)\\
&-\frac{1}{2}\left(\nabla_{\lambda}\nabla_{\mu}\left(F_1(\phi)\nabla^{\lambda}\phi\nabla_{\nu}\phi\right)+\nabla_{\lambda}\nabla_{\nu}\left(F_1(\phi)\nabla^{\lambda}\phi\nabla_{\mu}\phi\right)\right)\\
&+\frac{1}{2}\nabla_{\lambda}\nabla^{\lambda}\left(F_1(\phi)\nabla_{\mu}\phi\nabla_{\nu}\phi\right)+\frac{1}{2}g_{\mu\nu}\nabla_{\lambda}\nabla_{\gamma}\left(F_1(\phi)\nabla^{\lambda}\phi\nabla^{\gamma}\phi\right)\Big]
\end{aligned}
\ee
and 
\be\label{eq7a}
\begin{aligned}
T_{\mu\nu}^{GB}=&4\Big([\nabla_{\mu}\nabla_{\nu}F_2(\phi)]R-g_{\mu\nu}[\nabla_{\rho}\nabla^{\rho}F_2(\phi)]R-2[\nabla^{\rho}\nabla_{\mu}F_2(\phi)]R_{\nu\rho}-2[\nabla^{\rho}\nabla_{\nu}F_2(\phi)]R_{\nu\rho}\\
&+2[\nabla_{\rho}\nabla^{\rho}F_2(\phi)]R_{\mu\nu}+2g_{\mu\nu}[\nabla^{\rho}\nabla^{\sigma}F_2(\phi)]R_{\rho\sigma}-2[\nabla^{\rho}\nabla^{\sigma}F_2(\phi)]R_{\mu\rho\nu\sigma}\Big)
\end{aligned}
\ee
In this last expression the properties of the 4-dimensional GB invariant have been used (see \cite{sergei12}, \cite{farhoudi}).
Variating with respect to the scalar field gives the equation of motion
\be\label{eq7}
\begin{aligned}
&-\frac{1}{\sqrt{-g}}\partial_{\mu}\left[\sqrt{-g}\left(\xi R F_1(\phi)\partial^{\mu}\phi+\eta R^{\mu\nu}F_1(\phi)\partial_{\nu}\phi+\partial^{\mu}\phi\right)\right]+\frac{dV}{d\phi}+\\
&\frac{dF_1}{d\phi}\left(\xi R\partial_{\mu}\phi\partial^{\mu}\phi+\eta R_{\mu\nu}\partial^{\mu}\phi\partial^{\nu}\phi\right)-\frac{dF_2}{d\phi}{\cal G}=0
\end{aligned}
\ee
Considering the spatially-flat Friedmann-Robertson-Walker (FRW) metric,
\be\label{eq8}
ds^2=-dt^2+a(t)^2\left(dr^2+r^2d\Omega^2\right)
\ee
And assuming an homogeneous time-depending scalar field $\phi$ , the $(00)$ and $(11)$ components of the Eq. (\ref{eq2}), from (\ref{eq3}-\ref{eq7a}) take the form (with the Hubble parameter $H=\dot{a}/a$)
\be\label{eq9}
H^2=\frac{\kappa^2}{3}\rho_{eff}
\ee
with $\rho_{eff}$ given by
\be\label{eq9a}
\begin{aligned}
\rho_{eff}=&\Big[\frac{1}{2}\dot{\phi}^2+V(\phi)+9\xi H^2F_1(\phi)\dot{\phi}^2+3(2\xi+\eta)\dot{H}F_1(\phi)\dot{\phi}^2\\
&-3(2\xi+\eta)H F_1(\phi)\dot{\phi}\ddot{\phi}-\frac{3}{2}(2\xi+\eta)H \frac{dF_1}{d\phi}\dot{\phi}^3-24H^3\frac{dF_2}{d\phi}\dot{\phi}\Big]
\end{aligned}
\ee
and
\be\label{eq10}
-2\dot{H}-3H^2=\kappa^2 p_{eff}
\ee
with $p_{eff}$ given by
\be\label{eq10a}
\begin{aligned}
p_{eff}=&\Big[\frac{1}{2}\dot{\phi}^2-V(\phi)+3(\xi+\eta)H^2F_1(\phi)\dot{\phi}^2+2(\xi+\eta)\dot{H}F_1(\phi)\dot{\phi}^2\\
&+4(\xi+\eta)H F_1(\phi)\dot{\phi}\ddot{\phi}+2(\xi+\eta)H\frac{dF_1}{d\phi}\dot{\phi}^3\\
&+(2\xi+\eta)\left(F_1(\phi)\ddot{\phi}^2+F_1(\phi)\dot{\phi}\dddot{\phi}+\frac{5}{2}\frac{dF_1}{d\phi}\dot{\phi}^2\ddot{\phi}+\frac{1}{2}\frac{d^2F_1}{d\phi^2}\dot{\phi}^4\right)\\
&+8H^2\frac{dF_2}{d\phi}\ddot{\phi}+8H^2\frac{d^2F_2}{d\phi^2}\dot{\phi}^2+16H\dot{H}\frac{dF_2}{d\phi}\dot{\phi}+16H^3\frac{dF_2}{d\phi}\dot{\phi}\Big]
\end{aligned}
\ee
where we have assumed scalar field dominance (i.e. $T^m_{\mu\nu}=0$). The equation of motion for the scalar field (\ref{eq7}) takes the form
\be\label{eq11}
\begin{aligned}
&\ddot{\phi}+3H\dot{\phi}+\frac{dV}{d\phi}+3(2\xi+\eta)\ddot{H}F_1(\phi)\dot{\phi}+
3(14\xi+5\eta)H\dot{H}F_1(\phi)\dot{\phi}\\
&+\frac{3}{2}(2\xi+\eta)\dot{H}\left(2F_1(\phi)\ddot{\phi}+\frac{dF_1}{d\phi}\dot{\phi}^2\right)+
\frac{3}{2}(4\xi+\eta)H^2\left(2F_1(\phi)\ddot{\phi}+\frac{dF_1}{d\phi}\dot{\phi}^2\right)\\
&+9(4\xi+\eta)H^3F_1(\phi)\dot{\phi}-24\left(\dot{H}H^2+H^4\right)\frac{dF_2}{d\phi}=0
\end{aligned}
\ee
where the first three terms correspond to the minimally coupled field. In what follows we study the cosmological consequences of this equations, under some conditions that simplify the search for solutions.\\
\noindent An important simplification of the Eqs. (\ref{eq9}-\ref{eq11}) takes place, under the restriction on $\xi$ and $\eta$ given by 
\be\label{eq1a}
\eta+2\xi=0
\ee
Under this restriction all the kinetic couplings that appear in the action (\ref{eq1}) become summarized in the term $G_{\mu\nu}\partial^{\mu}\phi\partial^{\nu}\phi$, where $G_{\mu\nu}=R_{\mu\nu}-\frac{1}{2}g_{\mu\nu}R$. In this case the field equations (\ref{eq9}-\ref{eq11}) contain only second derivatives of the metric and the scalar field, avoiding problems with higher order derivatives \cite{capozziello1, sushkov}. The modified Friedmann equations (\ref{eq9}) and (\ref{eq10}) take the form
\be\label{eq12}
H^2=\frac{\kappa^2}{3}\left(\frac{1}{2}\dot{\phi}^2+V(\phi)+9\xi H^2F_1(\phi)\dot{\phi}^2-24H^3\frac{dF_2}{d\phi}\dot{\phi}\right)
\ee
and
\be\label{eq13}
\begin{aligned}
-2\dot{H}-3H^2=&\kappa^2\Big[\frac{1}{2}\dot{\phi}^2-V(\phi)-\xi\left(3H^2+2\dot{H}\right)F_1(\phi)\dot{\phi}^2-2\xi H\left(2F_1(\phi)\dot{\phi}\ddot{\phi}+\frac{dF_1}{d\phi}\dot{\phi}^3\right)\\
&+8H^2\frac{dF_2}{d\phi}\ddot{\phi}+8H^2\frac{d^2F_2}{d\phi^2}\dot{\phi}^2+16H\dot{H}\frac{dF_2}{d\phi}\dot{\phi}+16H^3\frac{dF_2}{d\phi}\dot{\phi}\Big]
\end{aligned}
\ee
The equation of motion reduces to
\be\label{eq14}
\begin{aligned}
&\ddot{\phi}+3H\dot{\phi}+\frac{dV}{d\phi}+3\xi H^2\left(2F(\phi)\ddot{\phi}+\frac{dF}{d\phi}\dot{\phi}^2\right)
+18\xi H^3F(\phi)\dot{\phi}+\\
&12\xi H\dot{H}F(\phi)\dot{\phi}-24\left(\dot{H}H^2+H^4\right)\frac{dF_2}{d\phi}=0
\end{aligned}
\ee
\noindent Note that, independently of $F_1(\phi)$, and for $F_2(\phi)=const.$, assuming an asymptotic behavior of the scalar field as $\phi=\phi_0=const.$, gives rise to de Sitter solution, as can be seen from Eqs. (\ref{eq9}) and (\ref{eq11}). From (\ref{eq9}) and (\ref{eq11}) it follows that $V=V_0=const$ and $H=H_0=\kappa\sqrt{V_0/3}$.\\
In the following sections we study cosmological solutions of Eqs. (\ref{eq12}) and (\ref{eq14}), giving rise to accelerated expansion and describing late time cosmological dynamics, according to current observations.
\section{Cosmological solutions in the time variable}
We start studying some solutions to Eqs. (\ref{eq12}) and (\ref{eq14}), in the important case when the scalar field potential is absent (i.e. $V=0$).  This leaves the two couplings $F_1(\phi)$ and $F_2(\phi)$ as the degrees of freedom that may characterize the cosmological dynamics. Predicting this couplings is very difficult task, but at least for the GB coupling there are some indications from fundamental theories like strings, that favor the exponential coupling (dilaton) \cite{bertolami}. Making $\dot{\phi}^2=\psi$, from Eq. (\ref{eq12}) with $V=0$, it follows
\be\label{eq14a}
F_1\psi=\frac{1}{3\xi}-\frac{\psi}{18\xi H^2}+\frac{8}{3\xi}H\frac{dF_2}{dt}
\ee
replacing in (\ref{eq14}) after simplifications, we obtain
\be\label{eq14b}
H\frac{d\psi}{dt} +6H^2\psi-\psi\frac{dH}{dt}+48H^3\frac{dH}{dt}\frac{dF_2}{dt}+72H^5\frac{dF_2}{dt}+24H^4\frac{d^2F_2}{dt^2}+12H^2\frac{dH}{dt}+18H^4=0
\ee
here we have set $\kappa^2=1$. Let's consider some solutions to this equation. In the following solutions we will consider the GB coupling
\be\label{eq14c}
\frac{dF_2(t)}{dt}=\frac{g}{H(t)}
\ee
which as will be seen, permits to find some interesting solutions. Replacing (\ref{eq14c}) in (\ref{eq14b}) one obtains
\be\label{eq14d}
H\frac{d\psi}{dt}+6H^2\psi-\psi\frac{dH}{dt}+12(2g+1)H^2\frac{dH}{dt}+18(4g+1)H^4=0
\ee
Bellow we study some viable late time cosmological solutions that can be extracted from this equation.\\
\noindent {\bf 1)} {\it Constant EoS}.\\
Here we consider the following behavior for the kinetic term
\be\label{eq15a}
\psi(t)=\lambda^2 H(t)^2
\ee
replacing in (\ref{eq14d}) one finds the solution
\be\label{eq15b}
H(t)=\frac{\eta}{t-C\eta},\,\,\,\,\,\, \eta=\frac{\lambda^2+24g+12}{6(\lambda^2+12g+3)}
\ee
where $C$ is the integration constant. This solution gives a constant effective equation of state (EoS) $w_{eff}=-1+\frac{2}{3\eta}$. If $\eta>0, C<0$, then $H(t)$ does not present time singularities, presenting power law behavior at late times. Nevertheless the EoS is constant all the time, and there is not transition deceleration-acceleration. If ($\eta<0, C<0$), then the solution describes pure phantom phase with Big Rip singularity at $t_0=\eta C$ \cite{sergei17}, \cite{cimento}. The scalar field can be found from (\ref{eq15a}) as
\be\label{eq15bb}
\phi=\lambda\eta\ln(t-C\eta)+\phi_0
\ee
this expression allows to write the time in terms of the scalar field as: $t=e^{\phi/(\lambda\eta)}+C\eta$ (making $\phi_0=0$). The GB coupling in terms of the scalar field is obtained using (\ref{eq14c})
\be\label{eq15bbb}
F_2(\phi)=\frac{g}{2\eta}e^{2\phi/(\lambda\eta)}+C(\frac{1}{\eta}-1)e^{\phi/(\lambda\eta)}
\ee
where we neglected the constant term which gives tribial (constant) coupling. The kinetic coupling is found from (\ref{eq14a}) 
\be\label{eq15bc}
F_1(\phi)=\frac{1}{3\xi}\left(\frac{1}{\lambda^2}-\frac{1}{6}\right)e^{2\phi/(\lambda\eta)}+\frac{8g}{3\xi}
\ee
so, if we take $\lambda^2=6$, then $F_1$ becomes constant given by $F_2=8g/(3\xi)$.\\
\noindent {\bf 2)} {\it Power law solution}.\\
In this known case the Hubble function is given by
\be\label{eq15c}
H(t)=\frac{p}{t}
\ee
which gives accelerated expansion provided $p>1$, and also has constant EoS ($w_{eff}=-1+2/(3p)$). Replacing in (\ref{eq14d}) one finds the solution
\be\label{eq15d}
\psi(t)=\frac{6p^2(2-3p-4g(3p-1))}{(6p-1)t^2}+C t^{-1-6p}
\ee
this expression takes a simpler form if we set the integration constant $C=0$ and $g=-3/4$: $\psi(t)=6p^2/t^2$. In this case the scalar field becomes 
\be\label{eq15e}
\phi=\sqrt{6}p\ln t+\phi_0
\ee
This solution allows to express the cosmological time $t$ in terms of the scalar field as $t=e^{\frac{\phi}{\sqrt{6}p}}$ (taking $\phi_0=0$). The GB coupling from (\ref{eq14c}) becomes
\be\label{eq15f}
F_2(\phi)=-\frac{3}{8p} e^{\frac{2\phi}{\sqrt{6}p}}
\ee
Note that although $F_2$ increases with time ($F_2=g t^2/(2p)$), the GB term decreases faster that $F_2$ increases, giving adequate behavior at late times. The kinetic coupling from (\ref{eq14a}) becomes constant, given by $F_1=-2/\xi$.\\
\noindent {\bf 3)} {\it Time dependent EoS I}.\\
An interesting solution which provides the transition from the decelerating universe to the accelerating one, is (see \cite{sergei11})
\be\label{eq15g}
H(t)=\gamma+\frac{p}{t}
\ee
with positive constants ($\gamma, p$). This Hubble function describes early time ($t\rightarrow 0$) power law behavior $a\propto t^p$ and late time dark energy dominated by the cosmological constant $H\sim \gamma$. The corresponding effective EoS is
\be\label{eq15gg}
w=-1+\frac{2p}{3(\gamma t+p)^2}
\ee
Here and in what follows, we use the symbol $w$ to mean $w_{eff}$. The transition from decelerated to accelerated expansion occurs at $t_c=(\sqrt{p}-p)/\gamma$. Note that in order for this transition to occur, $p$ should be in the interval $0<p<1$. For a given current EoS $w_0$ we can find a relation between $t_c$ and the current time $t_0$ (age of the universe). Using the above expression for $t_c$
and (\ref{eq15gg}) at $t_0$ we find
\be\label{eq15ggg}
t_c=\frac{3\sqrt{1+w_0}(1-\sqrt{p})}{\sqrt{6}-3\sqrt{1+w_0}\sqrt{p}}t_0
\ee
From this expression we can see that the closer to $-1$ $w_0$ is, the earlier the transition occurs.
Replacing in (\ref{eq14d}) and solving, one obtains the kinetic term $\psi$. But the solution takes simpler form if we choose $g=-3/4$, and set the integration constant equal to zero, which gives
\be\label{eq15h}
\psi(t)=\frac{6(p+\gamma t)^2}{t^2}
\ee
This equation can be easily integrated to find the scalar field (ignoring the integration constant)
\be\label{eq15i}
\phi(t)=\sqrt{6}\gamma t+\frac{\sqrt{6}p}{t}=\sqrt{6}\ln(t^p e^{\gamma t})
\ee
Solving this Eq. with respect to $t$, one finds: 
\be\label{eq15j}
t=\frac{p}{\gamma}W\left[(\gamma/p) e^{\phi/(\sqrt{6}p)}\right]
\ee
where $W$ is the Lambert $W$-function. The GB coupling, from (\ref{eq14c}) in terms of the scalar field is given by
\be\label{eq15k}
F_2(t)=\frac{3p}{4\gamma^2}\left(\ln\left[p\left(1+W\left[(\gamma/p) e^{\phi/(\sqrt{6}p)}\right]\right)\right]-W\left[(\gamma/p) e^{\phi/(\sqrt{6}p)}\right]\right)
\ee
and the kinetic coupling $F_1$ for this solution becomes constant $F_1=-2/\xi$.\\
\noindent {\bf 4)} {\it Time dependent EoS II}.\\
Let's consider the following evolutionary scenario
\be\label{eq15l}
H(t)=\frac{p}{t}+\frac{q}{t_s-t}
\ee
The first term guarantees the power law behavior at early times $a\propto t^p$, and the second term despite the fact that gives rise to Big Rip singularity at $t=t_s$, has the advantage that provides the crossing of the phantom barrier (see \cite{sergei11}). Therefore, this model presents three phases: early time decelerated ($p<1$), transition from decelerated to quintessence phase, and transition from quintessence to phantom phase. From (\ref{eq15l}) the scale factor is $a(t)=a_0 \frac{t^p}{(t_s-t)^q}$, and the EoS is given by
\be\label{eq15m}
w=-1+\frac{2}{3}\frac{t_sp(t_s-2t)+(p-q)t^2}{(t_s p+(q-p)t)^2}
\ee
we can obtain a relationship between the different critical times as follows. If we consider that at current time $t_0$ the EoS reaches the value $w_0=-1$ (which could be true according to observational data), then from (\ref{eq15m}) we find (in fact there are two roots, but we choose one)
\be\label{eq15n}
t_0=\frac{\sqrt{p}}{\sqrt{q}+\sqrt{p}}t_s
\ee
and the transition decelerated-accelerated expansion occurs at $t=t_c$, when $w_{c}=-1/3$. From (\ref{eq15m}) we find
\be\label{eq15o}
t_c=\frac{(p(p-q-1)+\sqrt{pq(q-p+1)})}{(q-p)(q-p+1)}t_s
\ee
from (\ref{eq15n}) and (\ref{eq15o}) we find the relation between the age of the universe $t_0$ and $t_c$
\be\label{eq15p}
\frac{t_0}{t_c}=\frac{(\sqrt{q}-\sqrt{p})(q-p+1)^{1/2}}{\sqrt{q}-\sqrt{p}(q-p+1)^{1/2}}
\ee
Thus, if we take $p=2/3$ which simulates matter dominance at early times, then for $q=1/8$ one finds $t_0/t_c\approx 1.6$, which is an appropriate relation according to observations. If we start from radiation dominance $p=1/2$, then for $q\approx 1/4$ one also finds $t_0/t_c\approx 1.6$. On the other hand, from (\ref{eq15n}) it follows that the Big Rip singularity occurs at the future and depends on the fraction $q/p$ (Big Rip singularities in modified $F(G)$ gravity have been studied in \cite{sergei18}). At this time the EoS reaches the value $w_{s}=-1-2/(3q)$. Hence, the unification of initial decelerated, accelerated and phantom universe may emerge in this model. Let's turn to the scalar model that gives rise to this scenario.\\
The kinetic term $\psi$ is obtained by replacing (\ref{eq15l}) in (\ref{eq14d}). The resulting equation has a simpler solution if we take $g=-3/4$, which is
\be\label{eq15pp}
\psi(t)=\frac{(q-p)t+pt_s}{t^{2+6p}(t_s-t)^2}\left[\left(6pt^{6p}+C(t_s-t)^{6q}t\right)t_s-\left(6(p-q)t^{6p}+C(t_s-t)^{6q}t\right)t\right]
\ee
One can make one more simplification if taking $C=0$, giving
\be\label{eq15ppp}
\psi(t)=6\frac{((q-p)t+pt_s)^2}{(t_s-t)^2t^2}
\ee
Considering this last case, we can explicitly integrate to obtain the scalar field as 
\be\label{eq15q}
\phi=\sqrt{6}\ln\left[\frac{t^p}{(t_s-t)^q}\right]
\ee
The GB coupling from (\ref{eq15l}) is
\be\label{eq15r}
F_2(t)=-\frac{3}{4}\left(\frac{qt_s}{(p-q)^2}t+\frac{t^2}{2(p-q)}+\frac{pqt_s^2}{(p-q)^3}\ln\left[p(t_s-t)+qt\right]\right)
\ee
and the kinetic coupling from (\ref{eq14a}) and (\ref{eq15ppp}) is found again as the constant $F_1=-2/\xi$.
In the special case $q=p$ (see \cite{sergei17}) we can express the GB coupling explicitly as function of the scalar field. In this case from (\ref{eq15q}) one can express the time as (setting $q=p$)
\be\label{eq15s}
t=\frac{t_s}{1+e^{-\phi/(\sqrt{6}p)}}
\ee
and integrating the Eq. (\ref{eq14c}) with $p=q$ one finds the GB coupling as
\be\label{eq15t}
F_2(\phi)=-\frac{t_s^2}{8p}\frac{e^{2\phi/(\sqrt{6}p)}(e^{\phi/(\sqrt{6}p)}+3)}{(1+e^{\phi/(\sqrt{6}p)})^3}
\ee
Note that the particular solution (\ref{eq15ppp}) for the kinetic term is actually $\psi=6H^2$.
\section{Cosmological solutions in the e-folding variable}
In this section we study some solutions to Eqs. (\ref{eq12}) and (\ref{eq14}) in the $x=\ln a$ variable. 
It is of interest to derive solutions that not only reproduce the current accelerated expansion according to observations, but that also have appropriate asymptotic behavior at early times. In the e-folding variable $x=\ln a$, we have $d/dt=H d/dx$. Then $\dot{\phi}^2=H^2(d\phi/dx)^2$ and the Eqs. (\ref{eq12}) and (\ref{eq14}) take the form
\be\label{eq15}
H^2=\frac{\kappa^2}{3}\left[\frac{1}{2}H^2\theta+V+9\xi H^4F_1\theta-24H^4\frac{dF_2}{dx}\right]
\ee
and 
\be\label{eq16}
\begin{aligned}
&\frac{1}{2}\frac{d}{dx}(H^2\theta)+3H^2\theta+\frac{dV}{dx}+9\xi H^2(F_1\theta)\frac{dH^2}{dx}+3\xi H^4\frac{d}{dx}(F_1\theta)\\
&+18\xi H^4(F_1\theta)-12H^2\frac{dH^2}{dx}\frac{dF_2}{dx}-24H^4\frac{dF_2}{dx}=0
\end{aligned}
\ee
where we have multiplied the Eq. (\ref{eq14}) by $\dot{\phi}$ and represented $(d\phi/dx)^2=\theta(x)$. Replacing the product $F_1\theta$ from (\ref{eq15}) into (\ref{eq16}) we arrive at the following equation
\be\label{eq17}
\begin{aligned}
&2H^4\frac{d\theta}{dx}+\theta H^2\frac{dH^2}{dx}+12H^4\theta+4H^2\frac{dV}{dx}-2V\frac{dH^2}{dx}-12H^2V+\frac{12}{\kappa^2}H^2\frac{dH^2}{dx}\\
&+\frac{36}{\kappa^2}H^4+72H^4\frac{dH^2}{dx}\frac{dF_2}{dx}+144H^6\frac{dF_2}{dx}+48H^6\frac{d^2F_2}{dx^2}=0
\end{aligned}
\ee
This is a first order differential equation for the functions $H$, $\theta$ and $V$, and of second order for the coupling $F_2$ (in fact is a first order for $dF_2/dx$). To have a solution we need to give three of this functions and integrate the equation with respect to the remaining one. It is desirable to have some definite criteria to choose the correct couplings and the potential in order to prove the predictive power of the theory.
There are some indications from the fundamental theories like superstring and supergravity that suggest potentials and GB coupling as exponentials of the scalar field, but strictly speaking, we still lack a definitive criterion for selecting potentials and couplings in the scalar tensor theory. Perhaps a good criterion to select the kinetic coupling as inverse of the squared scalar field ($F_1=1/\phi^2$) \cite{granda} is that the coupling constant $\xi$ becomes dimensionless, but though this is the simplest, is not the unique choice. As we have seen in the time variable, a simple exponential or even constant kinetic coupling, led to interesting cosmological solutions. Concerning the GB coupling, note that in (\ref{eq15bbb}) and (\ref{eq15f}) the GB coupling is expressed as exponential of the scalar field, while in (\ref{eq15k}) and (\ref{eq15t}) is expressed through exponentials of the scalar field. As we will see bellow, the choice of the GB coupling, allows to describe the cosmological dynamics compatible with  current observational data. 
To begin, we choose the GB coupling satisfying the restriction
\be\label{eq18}
\frac{dF_2}{dx}=\frac{g}{H^2}
\ee
where $g$ is a constant of dimension $L^{-2}$. This coupling lowers the power of $H$ by two and four in the last three terms of Eq. (\ref{eq17}), making it more tractable. Replacing (\ref{eq18}) in (\ref{eq17}) one obtains
\be\label{eq19}
\begin{aligned}
&2H^4\frac{d\theta}{dx}+\theta H^2\frac{dH^2}{dx}+12\theta H^4+4H^2\frac{dV}{dx}-2V\frac{dH^2}{dx}-12H^2V\\
&+\frac{12}{\kappa^2}H^2\frac{dH^2}{dx}+\frac{36}{\kappa^2}H^4+24g H^2\frac{dH^2}{dx}+144g H^4=0
\end{aligned}
\ee
Let's now consider different cosmological scenarios that we can extract from this equation (after solving this equation we can find the kinetic coupling $F_1$ from Eq. (\ref{eq15}). In this paper we consider an important class of solutions obtained in absence of potential.\\ In what follows we will introduce the scaled variables, but keeping the same symbols, namely we consider the Hubble function as the scaled one, i.e. $H\rightarrow H/H_0$,
the scaled potential energy $V\rightarrow \kappa^2V/H_0^2$, and the scaled kinetic term as will be clear bellow $\theta\rightarrow \kappa^2\theta$. Keeping the same symbols for these functions is also equivalent to set $\kappa^2=1$ and $H_0=1$, where $H_0$ is the current value of the Hubble parameter. After this rescalings and making $V=0$ in (\ref{eq19})
it follows
\be\label{eq20}
2H^2\frac{d\theta}{dx}+\theta\frac{dH^2}{dx}+12\theta H^2+12(1+2g)\frac{dH^2}{dx}+36(1+4g)H^2=0
\ee
This equation depends on two unknown functions, and we have to propose or constrain one of them in order to integrate the remaining one.
\subsection*{Solutions for a given $\theta(x)$}
\noindent {\bf 1)} {\it Constant kinetic term}.\\
Let's consider $\theta=const.=\lambda^2$. Replacing in (\ref{eq20}), we find
\be\label{eq21}
H^2=C e^{-2\sigma x},\,\,\,\,\,\ \sigma=\frac{6\left(12g+\lambda^2+3\right)}{24g+\lambda^2+12}
\ee
This solution corresponds to power law behavior $a(t)\propto t^{1/\sigma}$, describing presureless matter for $\sigma=3/2$, radiation for $\sigma=2$, and accelerated expansion provided $1/\sigma>1$. From the expression for $\theta$, it follows for the scalar field (up to additive constant): $\phi=\lambda x$. Integrating (\ref{eq18}), the GB coupling becomes
\be\label{eq22}
F_2(\phi)=\frac{g}{C}e^{2\sigma x}=\frac{g}{2\sigma C}e^{2\frac{\sigma}{\lambda}\phi}
\ee
here we neglected the constant of integration, since a constant coupled to GB term gives trivial results, due to the geometrical properties of the GB term.
From (\ref{eq15}) we find the expression for the coupling $F_1$
\be\label{eq22a}
F_1(\phi)=\frac{1}{3C\xi}\left(\frac{1+8g}{\lambda^2}-\frac{1}{6}\right)e^{\frac{2\sigma}{\lambda}\phi}
\ee
so, in absence of potential, exponential couplings give rise to power law behavior. \\
\noindent {\bf 2)} {\it Exponential behavior}.\\
Let's consider the exponential function for the kinetic term
\be\label{eq23}
\theta(x)=\lambda^2 e^{-\alpha x}
\ee
replacing in Eq. (\ref{eq20}) we find the following solution for the Hubble function
\be\label{eq24}
H(x)^2=\left(A e^{2\frac{(\alpha-6)}{\gamma}x}+B e^{(\alpha+2\frac{(\alpha-6)}{\gamma})x}\right)^{\gamma},\,\,\,\,\, \gamma=\frac{9-4g(\alpha-3)-2\alpha}{2g\alpha+\alpha}
\ee
An interesting case of this solution takes place when $\gamma=1$, giving
\be\label{eq25}
H(x)^2=A e^{2(\alpha-6)x}+B e^{(3\alpha-12)x}
\ee
taking $\alpha=4$ gives a solution that describes two asymptotically important cases: radiation dominance at high redshifts ($x\rightarrow -\infty$) and cosmological constant at far future. The value $\alpha=9/2$ describes matter dominance at high redshift and phantom behavior (negative power law) at far future (if $\gamma=1$). In this case if we take $A=0.3$, then the transition decelerated-accelerated expansion occurs at $z_t\sim 0.59$, with current EoS $w_0=-1.05$, presenting quintom behavior (see bellow). The equation of state that follows from solution (\ref{eq24}) with the flatness condition $A+B=1$, is (in the redshift variable, using $1+z=e^{-x}$)
\be\label{eq26}
w(z)=3-\frac{2}{3}\alpha-\frac{(1-A)\alpha\gamma}{3\left(1+A[(1+z)^{\alpha}-1]\right)}
\ee
the important limits of the EoS are for positive $\alpha$. The asymptotical EoS at early times is
\be\label{eq27}
\lim_{z\to \infty}w(z)=w_{\infty}=3-\frac{2}{3}\alpha
\ee
while
\be\label{eq28}
\lim_{z\to -1}w(z)=w_{-1}=3-\frac{2}{3}\alpha-\frac{\alpha\gamma}{3}
\ee
describes the asymptotic future value of the EoS. The current EoS is
\be\label{eq29}
w(z=0)=w_0=3-\frac{2}{3}\alpha-\frac{1}{3}(1-A)\alpha\gamma
\ee
For a given set of parameters ($\alpha, \gamma, A$), if the solution starts in a decelerating regime, the transition deceleration-acceleration occurs at the redshift
\be\label{eq29a}
z_t=-1+(1-A)^{1/\alpha}\left(\frac{10-2\alpha-\alpha\gamma}{2A(\alpha-5)}\right)^{1/\alpha}
\ee
Thus for instance, we can reach similar results as obtained with the $\Lambda$CDM model: if we take $\alpha=9/2$, then $w_{\infty}=0$, describing the matter dominance era, and taking
$\gamma=2/3$ the universe evolves towards cosmological constant phase (de Sitter) $w_{-1}=-1$. Taking $\gamma=1$ gives $w_{-1}=-3/2$ and the solution evolves from matter dominance, passing through quintessence, to finally enter in the  phantom phase, presenting quintom behavior. Note that in principle, using Eqs. (\ref{eq27})-(\ref{eq29}), we can satisfy any given in advance asymptotical ($w_{\infty},w_{-1}$) and current ($w_0$) behavior, within observational expectations. If we remove the GB coupling (i.e. $g=0$), then the power $\gamma$ can not be chosen independently of $\alpha$, which makes the solution not interesting.\\
Integrating in (\ref{eq23}) one finds the scalar field
\be\label{eq29b}
\phi=\frac{2\lambda}{\alpha}e^{-\alpha x/2}+\phi_0
\ee
and from (\ref{eq18}), the GB coupling is
\be\label{eq29c}
F_2(\phi)=\frac{g}{2(6-\alpha)A^{\gamma}}\phi^{\frac{4(\alpha-6)}{\alpha}}\hspace{0.1cm} _{2}F_1\left[-2+\frac{12}{\alpha},\gamma,-1+\frac{12}{\alpha},-\frac{B}{A\phi^2}\right]
\ee
where we have used (\ref{eq29b}), setting $\phi_0=0$, and rescaled $\phi\rightarrow \frac{\alpha}{2\lambda}\phi$. 
The kinetic coupling is found from Eq. (\ref{eq15}) as
\be\label{eq29d}
F_1(\phi)=\frac{\phi^{\frac{4(\alpha-6)}{\alpha}+2\gamma}}{3\xi\left(A\phi^2+B\right)^{\gamma}}\left(\frac{1+8g}{\lambda^2\phi^2}-\frac{1}{6}\right)
\ee
\subsection*{Solutions for a given $H(x)$}
The cosmological equations can also be consistently solved by a given behavior of Hubble function (or the scale parameter). In this case we reconstruct the scalar field and the couplings. Let's consider some important solutions\\

\noindent {\bf 1)} {\it $\Lambda$CDM-type solution}.\\
The Hubble function
\be\label{eq30}
H^2=\Omega_1 e^{-\alpha x}+\Omega_2 
\ee
reproduces the power law behavior at early times, describing radiation or matter dominance, and late time behavior governed by the cosmological constant. It can also serves as purely dark energy model for appropriate values of $\alpha$. Replacing in Eq. (\ref{eq19}) and solving with respect to $\theta$ one obtains 
\be\label{eq31}
\begin{aligned}
\theta(x)=&C\left(\Omega_1+\Omega_2 e^{\alpha x}\right)^{-1/2}e^{(\alpha-12)x/2}-3(1+4g)+\\
&\frac{3(3+4g)\alpha\sqrt{\Omega_1}}{12-\alpha}\left(\Omega_1+\Omega_2 e^{\alpha x}\right)^{-1/2}\hspace{0.1cm} _{2}F_1\left[\frac{1}{2},-\frac{1}{2}+\frac{6}{\alpha},\frac{1}{2}+\frac{6}{\alpha},-\frac{\Omega_2 e^{\alpha x}}{\Omega_1}\right]
\end{aligned}
\ee
this solution is considerably simplified for $g=-3/4$, giving
\be\label{eq32}
\theta(x)=C\left(\Omega_1+\Omega_2 e^{\alpha x}\right)^{-1/2}e^{(\alpha-12)x/2}+6
\ee
and even simpler if we take the integration constant $C=0$. In this case $\theta(x)=6$, and after integration the scalar field is obtained as $\phi=\sqrt{6}x+\phi_0$. In fact is easy to check directly, that this particular value $\theta=6$ satisfies the Eq. (\ref{eq19}) with $H$ given by (\ref{eq29}).\\
For this particular solution, the scalar field is given by $\phi=\sqrt{6}x+\phi_0$, and from (\ref{eq18}) and (\ref{eq30}) we may write the GB coupling explicitly in terms of the scalar field, as
\be\label{eq33}
F_2(\phi)=-\frac{3}{4\alpha\Omega_2}\ln\left[\Omega_1+\Omega_2 e^{\alpha\phi/\sqrt{6}}\right]
\ee
from Eq. (\ref{eq15}) we find the kinetic coupling
\be\label{eq34}
F_1(\phi)=-\frac{1}{3\xi}\frac{e^{\alpha\phi/\sqrt{6}}}{\Omega_1+\Omega_2 e^{\alpha\phi/\sqrt{6}}}
\ee
Therefore, even without the cosmological constant nor cold dark matter in the action, the cosmology of the $\Lambda$CDM model can be reproduced by the combined effect of the kinetic and GB coupling.\\
\noindent {\bf 2)} {\it The Chaplygin gas solution}.\\
The Hubble function for the Chaplygin gas has the known form \cite{kamenshchick} (setting $\kappa^2=1$ and with appropriate normalization of $A$ and $B$)
\be\label{eq35}
H^2=\left[A+B e^{-6x}\right]^{1/2}
\ee
This solution comes from the EoS obeyed by the Chaplygin gas, $p=-\frac{A}{\rho}$. And has the known properties of describing the presureless matter dominance epoch of the universe at early times ($a<<1$, normalizing the current value of $a$ to $1$), and the future universe dominated by the cosmological constant, evolving towards de Sitter phase at $a>>1$.\\
Replacing (\ref{eq35}) in (\ref{eq19}), we arrive at the equation
\be\label{eq36}
2\left(A+Be^{-6x}\right)\frac{d\theta}{dx}+3\left(4A+3B e^{-6x}\right)\theta(x)+36\left(A+4A g+2Bg e^{-6x}\right)=0
\ee
which has the solution
\be\label{eq37}
\begin{aligned}
\theta(x)=&-\frac{1}{A+B e^{-6x}}\left(3A+12Ag+3B(1+4g)e^{-6x}+Ce^{-6x}(A+B e^{-6x})^{3/4}\right)\\
&+\frac{B^{1/4}e^{-3x/2}}{(A+B e^{-6x})^{1/4}}(3+4g)\hspace{0.1cm} _{2}F_1\left[\frac{3}{4},\frac{3}{4},\frac{7}{4},-\frac{A e^{6 x}}{B}\right]
\end{aligned}
\ee
This solution substantially simplifies under the restriction $g=-3/4$, becoming
\be\label{eq38}
\theta(x)=\frac{C e^{-9x/2}}{(B+A e^{6x})^{1/4}}+6
\ee
If we choose the integration constant $C=0$, then $\theta(x)=6$, and after elementary integration gives $\phi(x)=\sqrt{6}x+\phi_0$. For this particular solution, we find the GB coupling from (\ref{eq18}) and (\ref{eq35}) as (up to additive constant)
\be\label{eq39}
F_2(\phi)=-\frac{1}{4\sqrt{A}}\ln\left[A+\sqrt{A}\left(A+ Be^{-\sqrt{6}\phi}\right)^{1/2}\right]
\ee
and the kinetic coupling
\be\label{eq40}
F_1(\phi)=-\frac{1}{3\xi}\frac{e^{\sqrt{6}\phi}}{\left(B+A e^{\sqrt{6}\phi}\right)^{1/2}}
\ee
where we have set $\phi_0=0$. Thus, there is a particularly simple solution $\phi=\sqrt{6}x+\phi_0$, giving rise to Chaplygin gas cosmology with GB and kinetic couplings as simple functions of exponentials of the scalar field. Is worth to mention that this solution was obtained in absence of potential, which saves us having to put by hand one more input in the model.\\
\noindent {\bf 3)} {\it The generalized Chaplygin gas solution}.\\
The generalization of the Chaplygin gas has also been considered to describe in a unified way the dark matter and dark energy. It's equation of state is given by
\be\label{eq41}
p=-\frac{A}{\rho^{\alpha}} 
\ee
where $\rho$ and $p$ are the energy density and pressure of the generalized Chaplygin gas, $A$ is a positive constant and $\alpha$ is considered to lie in the range $0<\alpha\leq 1$, which guarantees the stability and causality \cite{bilic}. The value $\alpha=1$ gives the original Chaplygin gas. The EoS (\ref{eq41}) has an equivalent field theory representation in a generalization of the Born-Infeld theory \cite{bento1}. In the scalar field representation used in \cite{bento1}, the Born-Infeld Lagrangian density is reproduced for $\alpha=1$. Solving the continuity equation and appropriately rescaling the constant $A$ and the integration constant $B$ (setting $\kappa^2=1)$, leads to the Hubble function 
\be\label{eq42}
H^2(x)=\left[A+B e^{-3(\alpha+1)x}\right]^{\frac{1}{1+\alpha}}
\ee
replacing this function in (\ref{eq20}) gives
\be\label{eq43}
2\left(B+A e^{3(1+\alpha)x}\right)\frac{d\theta}{dx}+3\left(3B+4Ae^{3(1+\alpha)x}\right)\theta(x)+36\left(2gB+(1+4g)Ae^{3(1+\alpha)x}\right)=0
\ee
Integrating this equation gives the following kinetic term
\be\label{eq44}
\begin{aligned}
\theta(x)=&\frac{Ce^{-9x/2}}{(B+Ae^{3(1+\alpha)x})^{\frac{1}{2(1+\alpha)}}}\\
&-8gB^{\frac{1}{2(1+\alpha}}\left(B+A e^{3(1+\alpha)x}\right)^{-\frac{1}{2(1+\alpha}}\hspace{0.1cm} _{2}F_1\left[1-\frac{1}{2(1+\alpha)},\frac{3}{2(1+\alpha)},\frac{5+2\alpha}{2(1+\alpha)},-\frac{A e^{3(1+\alpha) x}}{B}\right]\\
&-\frac{12(1+4g)A}{(5+2\alpha)B^{\frac{1+2\alpha}{2+2\alpha}}}e^{3(1+\alpha)x}\left(B+A e^{3(1+\alpha)x}\right)^{-\frac{1}{2(1+\alpha)}}\times\\
&\hspace{0.1cm} _{2}F_1\left[1-\frac{1}{2(1+\alpha)},1+\frac{3}{2(1+\alpha)},2+\frac{3}{2(1+\alpha)},-\frac{A e^{3(1+\alpha) x}}{B}\right]
\end{aligned}
\ee
this expression drastically simplifies if $g=-3/4$
\be\label{eq45}
\theta(x)=\frac{Ce^{-9x/2}}{(B+Ae^{3(\alpha+1)x})^{\frac{1}{2(1+\alpha)}}}+6
\ee
Is remarkable that, again the simplest particular solution to Eq. (\ref{eq43}) with $g=-3/4$, is $\theta(x)=6$. In this case the GB coupling, after integration of (\ref{eq18}) is
\be\label{eq46}
F_2(\phi)=-\frac{1}{4B^{\frac{1}{1+\alpha}}}e^{3\phi/\sqrt{6}}\hspace{0.1cm} _{2}F_1\left[\frac{1}{1+\alpha},\frac{1}{1+\alpha},\frac{2+\alpha}{1+\alpha},-\frac{A e^{3(1+\alpha)\phi/\sqrt{6}}}{B}\right]
\ee
and the kinetic coupling from (\ref{eq15}) is
\be\label{eq47}
F_1(\phi)=-\frac{1}{3\xi}\left[A+B e^{-\frac{3(1+\alpha)}{\sqrt{6}}\phi}\right]^{-\frac{1}{1+\alpha}}
\ee
where we used the expression for the scalar field with $\phi_0=0$. The parameters $A$ and $B$ can be fixed according to the restrictions imposed by observational data on the Chaplygin gas models. Note that in all solutions (\ref{eq30}), (\ref{eq35}) and (\ref{eq42}), the universe evolves towards de Sitter phase. Is worth to note that the present reconstruction was made without appeal to the potential term, which is an advantage provided by the GB coupling. A reconstruction of the Chaplygin and Generalized Chaplygin gas cosmologies in scalar field model with kinetic couplings to curvature, have been presented in \cite{granda3}\\
\section{Discussion}
We have studied late time dark energy solutions in a scalar field model with kinetic terms coupled to curvature and a Gauss Bonnet term coupled to scalar field. The GB coupling has the advantage that does not make contributions higher than second order (in the metric) to the equations of motion, and therefore does not introduce ghost terms into the theory. Given a constant or exponential (of the e-folding variable) kinetic term, we have obtained explicit solutions for the Hubble function, and particularly a new viable solution (\ref{eq24}), leading to different equations of state at different epochs, including decelerated, accelerated and phantom regimes. A remarkable aspect of the solution (\ref{eq24}) is that we can satisfy any given in advance asymptotic and current behavior of the EoS. The model can also be reconstructed for known cosmological solutions describing early time radiation (matter) dominance with transition to accelerated universe with dark energy dominance, and in some cases even with one more transition to phantom phase. This means that we can think the energy-momentum tensor as being made of two components: one behaving like a presureless matter and other like dark energy with negative pressure. Nevertheless, concerning the presureless matter component, in order to be consistent with observations, we still need to investigate the clustering properties of the scalar field at scales of the order of the cosmic structures we observe in the universe.\\
In the e-folding variable we reconstructed the model for a class of important solutions, namely for the $\Lambda$CDM-type solution, and for the known Chaplygin and generalized Chaplygin gas cosmological solutions. An interesting aspect of these solutions is that for the specific numbers $g=-3/4$ and $\theta=6$, the expressions for the scalar field and the couplings have been considerably simplified. It is remarkable that in all cosmological scenarios considered, the potential term is absent, which saves us having to either implement a restriction on the potential or propose an ad hoc potential. Nevertheless, the presence of potential would bring us more possible solutions and will be considered elsewhere.\\
We have shown that it is possible to construct viable solutions to the dark energy problem with different asymptotic behaviors, in the context of the scalar field model with kinetic and GB couplings contributing to the total energy density. All obtained solutions for the scalar field and couplings $F_1$ and $F_2$, are simple enough to attract the attention on this model. Hence, the present model extends the number of possible evolutionary scenarios that explain the nature of the dark energy.  

\end{document}